\documentclass{jetpl}
\twocolumn

\rus

\title{Effect of correlations and doping on the spin susceptibility of iron pnictides: the case of KFe$_{2}$As$_{2}$}

\rtitle{Effect of correlations and doping on the spin susceptibility of iron pnictides: the case of KFe$_{2}$As$_{2}$}

\sodtitle{}

\author{S.\,L.~Skornyakov$^{+,*}$, V.\,I.~Anisimov$^{+,*}$, and D.~Vollhardt$^{\times}$}

\rauthor{S.\,L.~Skornyakov, V.\,I.~Anisimov, D.~Vollhardt}

\sodauthor{}

\address{$^{+}$Institute of Metal Physics, Russian Academy of Sciences,
S. Kovalevskaya Str. 18, 620990 Yekaterinburg, Russia \\
$^{*}$Ural Federal University, 620002 Yekaterinburg, Russia\\
$^{\times}$ Theoretical Physics III, Center for Electronic
Correlations and Magnetism, Institute of Physics, University of
Augsburg, D-86135 Augsburg, Germany}

\dates {\today}{*}

\abstract{
The temperature dependence of the paramagnetic susceptibility of the iron 
pnictide superconductor KFe$_{2}$As$_{2}$ and its connection with the 
spectral properties of that material is investigated by a combination of 
density functional theory (DFT) in the local density approximation and 
dynamical mean-field theory (DMFT). Unlike other iron pnictide parent 
compounds where the typical oxidation state of iron is 2, the formal 
valence of Fe in KFe$_{2}$As$_{2}$ is 2.5, corresponding to an effective 
doping with 0.5 hole per iron atom compared to, for example, BaFe$_{2}$As$_{2}$. 
This shifts the chemical potential and thereby reduces the distance between 
the peaks in the spectral functions of KFe$_{2}$As$_{2}$ and the Fermi 
energy as compared to BaFe$_{2}$As$_{2}$. The shift, which is clearly seen 
on the level of DFT as well as in DMFT, is further enhanced by the strong 
electronic correlations in KFe$_{2}$As$_{2}$. In BaFe$_{2}$As$_{2}$ the 
presence of these peaks results [Phys. Rev. B {\bf 86}, 125124 (2012)] 
in a temperature increase of the susceptibility up to a maximum at $\sim 1000$~K. 
While the temperature increase was observed experimentally the decrease 
at even higher temperatures is outside the range of experimental observability. 
We show that in KFe$_{2}$As$_{2}$ the situation is different. Namely, 
the reduction of the distance between the peaks and the Fermi level due 
to doping shifts the maximum in the susceptibility to much lower temperatures, 
such that the decrease of the susceptibility becomes visible in experiment.
}


\begin{document}

\maketitle


{\bf Introduction.}~
The discovery of high-temperature superconductivity in fluorine-doped 
LaFeAsO in 2008 by Kamihara et~al.~\cite{Kamihara06} has placed the 
iron-arsenic systems into the center of activity of solid state physics. 
To date a variety of Fe-based superconductors have been found. The so-called 
'122' family (compounds with the common formula {\it AE}Fe$_{2}$As$_{2}$, 
where {\it AE}=alkaline earth element) is the most studied one. Unlike other 
iron-arsenic systems, undoped compounds of the '122' family superconduct 
under pressure, with $T_c$ up to 29~K~\cite{Kreyssig, Alireza}. The highest 
critical temperature, $T_{c}=38$~K, in this family is detected in the 
potassium doped compound Ba$_{1-x}$K$_{x}$Fe$_{2}$As$_{2}$ with $x=0.4$
~\cite{Rotter,Chen,Sasmal}. Therefore KFe$_{2}$As$_{2}$ is the end member 
of the family and can be considered a parent compound in which superconductivity 
emerges under chemical doping. Although the critical temperature is rather 
low in KFe$_{2}$As$_{2}$ ($T_c$=3.8~K~\cite{Kihou}) this material is a rare 
example of a stoichiometric pnictide superconductor.

It is widely accepted that Coulomb correlations are crucial for the 
understanding of many aspects of the physics of pnictides~\cite{Aichhorn, 
SkornyakovLaFeAsO, SkornyakovBaFe2As2, SkornyakovLaFePO}. Correlation effects 
in KFe$_{2}$As$_{2}$ were intensively studied: Terashima et~al.~\cite{Terashima} 
performed de Haas-van Alphen measurements of the Fermi surface in 
KFe$_{2}$As$_{2}$. They detected unusually large effective mass renormalizations 
and big differences in the masses of different bands, which is not found in 
other pnictides. The enhancement of the band mass was also measured by 
Yoshida et~al.~\cite{Yoshida} in angular-resolved photoemission spectroscopy 
(ARPES) experiments. Hardy et~al.~\cite{Hardy} employed the Gutzwiller 
slave-boson mean-field method to study the strength of Coulomb correlations 
in KFe$_{2}$As$_{2}$. They confirmed the experimental conclusions of Terashima 
et~al.~\cite{Terashima} and Yoshida et~al.~\cite{Yoshida}, and proposed 
an orbital-selective scenario for its spectral properties.    

A characteristic feature of the magnetic properties observed in the '1111' 
(compounds like LaFeAsO) and '122' pnictide classes is the unusual linear-temperature 
increase of the paramagnetic susceptibility~\cite{Wang,Klingeler}. There are two 
explanations of this phenomenon based either on the assumption of strong 
antiferromagnetic fluctuations in a two-dimensional Fermi liquid~\cite{Korshunov}, 
or of peculiarities of the single-particle spectra~\cite{SkornyakovLaFeAsO,
SkornyakovBaFe2As2}. The temperature increase of the susceptibility was considered 
a universal property~\cite{Zhang1} of the pnictide superconductors and parent 
systems. By contrast, Cheng et~al.~\cite{Cheng} reported that in KFe$_{2}$As$_{2}$ 
the magnetic susceptibility increases only at quite low temperatures, i.e., 
below 100~K, and then decreases slowly at least up to 300~K. The origin of 
that temperature dependence of the magnetic susceptibility of KFe$_{2}$As$_{2}$ 
and its connection with the magnetic properties of the other end member of the 
'122' family, BaFe$_{2}$As$_{2}$, had not been explained yet. In particular, 
it was not studied by first-principle methods.       

Today the most powerful technique that can account for correlation effects in 
real compounds and can describe the physics of the correlated paramagnetic phase, 
is the LDA+DMFT approach~\cite{LDADMFT}. This method combines the advantage of 
density-functional theory (typically in the local density approximation (LDA)) 
to describe the material-specific electronic structure of a weakly correlated 
system, with the ability of the dynamical mean-field theory~\cite{DMFT} to treat 
the complete range of Coulomb correlations between the electrons in partially 
filled shells. 

In this work, we investigate the temperature evolution of the paramagnetic 
susceptibility in KFe$_{2}$As$_{2}$ in the framework of LDA+DMFT. We compare 
our results with experiments and our previously published LDA+DMFT 
data~\cite{SkornyakovBaFe2As2} obtained for the isostructural compound 
BaFe$_{2}$As$_{2}$. Thereby we demonstrate that the mechanism explaining the 
anomalous temperature behavior of the magnetic susceptibility in iron pnictides 
proposed in our previous investigations~\cite{SkornyakovLaFeAsO, SkornyakovBaFe2As2} 
also allows one to understand the difference between the magnetic properties 
of these compounds and those of KFe$_{2}$As$_{2}$. 


{\bf Technical details.}~  
In the LDA+DMFT formalism employed here the material-specific band dispersion 
obtained within LDA is used as a starting point. Then matrix elements of the 
effective Hamiltonian $H^{\mathrm{LDA}}({\bf k})$ are computed in the subspace 
of Wannier functions with the symmetry of $p$ and $d$ states using the projection 
procedure~\cite{projection}. In the second step, the Coulomb interaction matrix 
$U^{\sigma\sigma^{\prime}}_{mm^{\prime}}$, parametrized by Slater integrals 
$F_{0}$, $F_{2}$ and $F_{4}$, is calculated for each atom with partially filled 
shells. The values of $F_{0}$, $F_{2}$, $F_{4}$ are computed using the on-site 
effective Coulomb parameter $U$ and intraatomic exchange parameter $J$. Finally, 
the following many-electron Hamiltonian is iteratively solved by DMFT on the 
Matsubara contour:
\begin{eqnarray}
\hat H({\bf k})=\sum_{{\bf k},im,jm^{\prime},\sigma}
(H^{\mathrm{LDA}}_{im,jm^{\prime}}({\bf k})-H^{\mathrm{DC}}_{im,jm^{\prime}})
\hat a^{\dagger}_{{\bf k},im\sigma}\hat a_{{\bf k},jm^{\prime}\sigma}
\\
\nonumber
+
\frac{1}{2}\sum_{i,(m\sigma)\ne(m^{\prime}\sigma^{\prime})} 
U^{\sigma\sigma^{\prime}}_{mm^{\prime}}
\hat n_{im\sigma}^{d}\hat n_{im^{\prime}\sigma^{\prime}}^{d}.
\end{eqnarray}
Here $\hat a^{\dagger}_{{\bf k},im\sigma}$ is the Fourier transform of 
$\hat a^{\dagger}_{im\sigma}$ which creates an electron on the atom $i$ 
in the state $\left| m\sigma\right>$, where $m$ labels the orbitals and 
$\sigma=\uparrow,\downarrow$ corresponds to the spin projection. The 
particle number operator $\hat n_{im\sigma}^{d}$ acts on the states 
localized at the atoms with partially filled shells (Fe-$d$ states in 
the present study). The term $H^{\mathrm{DC}}$ stands for a double-counting 
correction which corresponds to the Coulomb interaction energy already 
accounted for by LDA (see below).

In the present work the LDA band structure is calculated with the ELK 
full-potential code~\cite{elk} with default parameters of the LAPW basis. 
By employing the constrained LDA method~\cite{constr} we obtained the 
interaction parameters $U=3.5$~eV and $J=0.85$~eV. These values are 
typical for the pnictides and are in good agreement with previous 
estimations~\cite{Aichhorn,LaFeAsO_JPCM}. The DMFT auxiliary impurity 
problem was solved by the hybridization function expansion quantum 
Monte-Carlo method~\cite{ctqmc}. The double-counting term is a diagonal 
matrix with only nonzero elements in the $d-d$ block expressed in the 
form $E^{\mathrm{DC}}=\bar U(n_{d}-0.5)$, where $n_{d}$ is the number 
of Fe-$d$ electrons calculated within LDA+DMFT and $\bar U$ is the 
average Coulomb parameter for the $d$ states. This form of $H^{\mathrm{DC}}$ 
yields reliable results for magnetic and spectral properties of iron 
pnictides~\cite{SkornyakovLaFeAsO, SkornyakovBaFe2As2,SkornyakovLaFePO,
Aichhorn}. 

The orbitally-resolved spectral functions $A_{i}(\omega)$ were computed as 
the diagonal elements of the real-energy Green function
\begin{equation}
A_{i}(\omega)=\sum_{\bf k}
[
I(\omega+\mu)-(H^{\mathrm{LDA}}({\bf k})-H^{\mathrm{DC}})-\Sigma(\omega)
]_{ii}^{-1},
\end{equation}
where $\mu$ is the chemical potential calculated within DMFT, $\Sigma(\omega)$ 
is the self-energy obtained with the use of Pad\' e approximants~\cite{Pade}, 
and $I$ is the identity matrix.

The uniform magnetic susceptibility $\chi(T)$ was calculated as the response 
to a small external magnetic field,
\begin{equation}
\chi(T)=\frac{\Delta M(T)}{\Delta E},
\end{equation}
where $\Delta E$ is the energy correction corresponding to the field and 
$\Delta M=|N_{\uparrow}(T)-N_{\downarrow}(T)|$ is the occupation difference 
between the spin projections.  


\begin{figure}[t]
\centering \vspace{0.0mm}
\includegraphics[width=0.85\linewidth,angle=0]{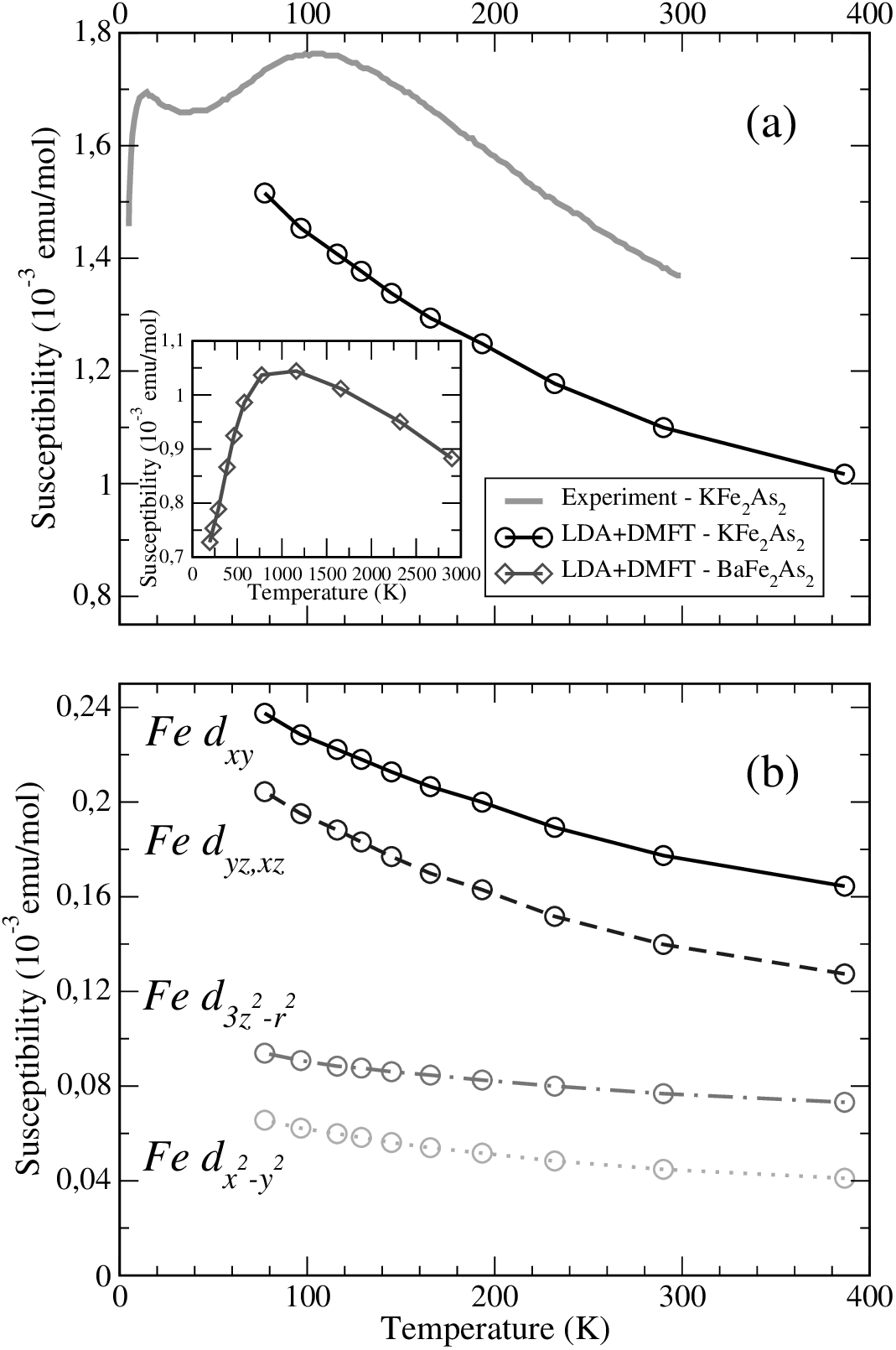}
\vspace{0.0mm}
\caption{Fig.1.~Temperature dependence of the uniform susceptibility 
of KFe$_{2}$As$_{2}$. (a) The static susceptibility as a function of 
temperature computed within LDA+DMFT (circles) is shown in comparison 
with experimental data of Cheng et~al.~\cite{Cheng} (solid curve). 
The inset shows the LDA+DMFT result for the susceptibility of BaFe$_{2}$As$_{2}$ 
from Ref.~\cite{SkornyakovBaFe2As2}. (b) Temperature behavior of the 
orbitally-resolved contributions to the total susceptibility calculated 
by LDA+DMFT}
\label{DMFTsuscvsExp}
\end{figure}
{\bf Temperature dependence of the uniform magnetic susceptibility.}~
In the upper panel of Fig.\ref{DMFTsuscvsExp} the temperature behavior 
of the static magnetic susceptibility of KFe$_{2}$As$_{2}$ as computed 
within LDA+DMFT is shown in comparison with the experimental result of 
Cheng et~al.~\cite{Cheng}. Both experimental and theoretical curves 
show a monotonic decrease in the temperature interval from 125 to 300~K. 
The slope of the calculated curve is in good agreement with experiment, 
while its absolute value is by about 20\verb\%\ smaller. The maximum in 
the experimental susceptibility observed at 100~K is not reproduced in 
the calculation. Temperatures lower than 77~K are not accessible in the 
present study. The temperature dependence of the LDA+DMFT calculated 
paramagnetic susceptibility in the other end member of the '122' family 
is shown in the inset of Fig.\ref{DMFTsuscvsExp}. The similarities and 
differences of the curves are discussed in the Discussion section. The 
orbitally resolved Fe-$d$ contributions to the total susceptibility are 
presented in the lower panel of Fig.\ref{DMFTsuscvsExp}. The susceptibilities 
corresponding to the Fe-$d$ orbitals all show a decreasing behavior with 
temperature. The largest contributions come from the $xy$ and $yz(xz)$ 
orbitals. 

\begin{figure}[t]
\centering \vspace{0.0mm}
\includegraphics[width=0.85\linewidth,angle=0]{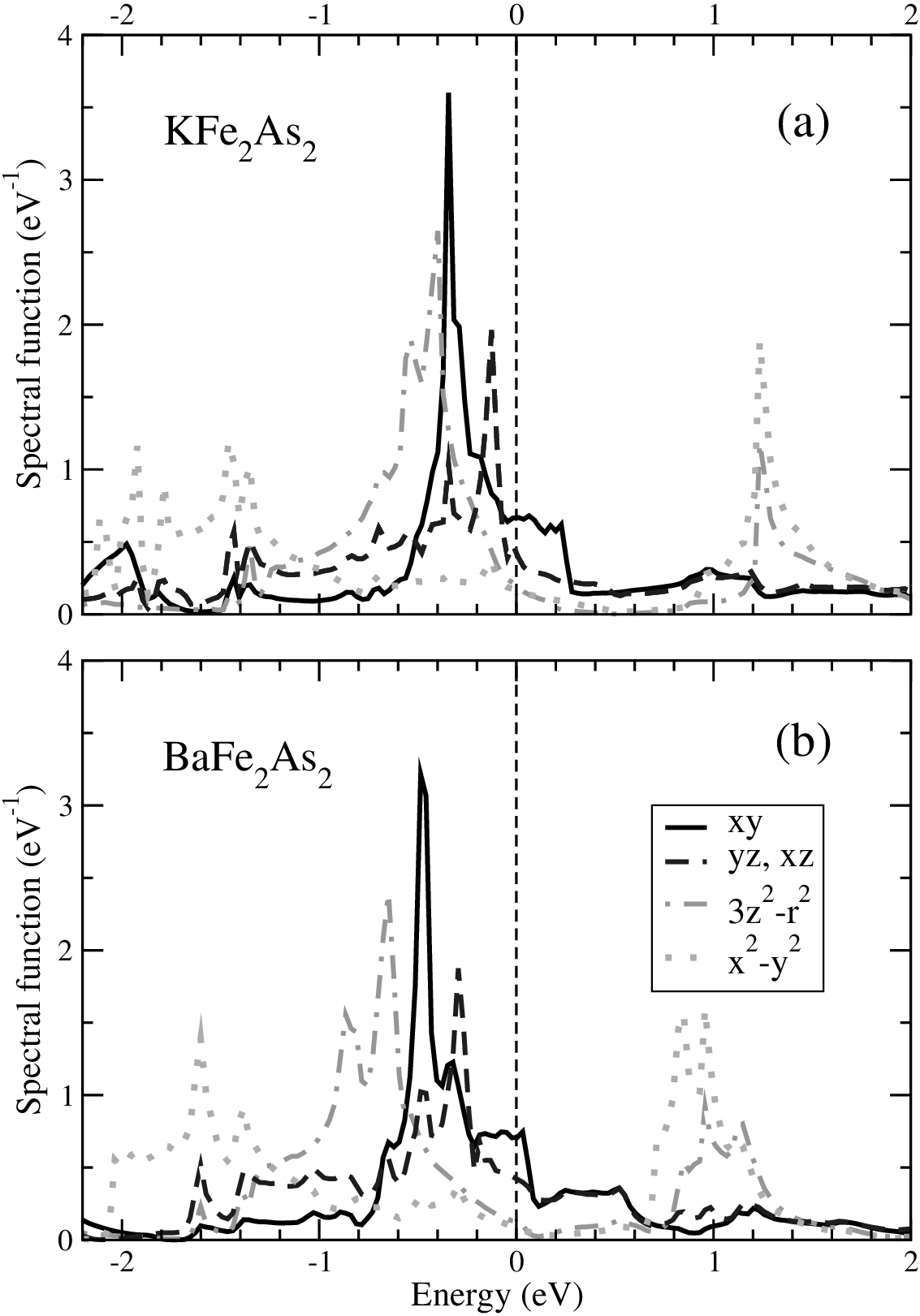}
\vspace{0.0mm}
\caption{Fig.2.~Fe-{\it d} orbitally-resolved densities of states of 
KFe$_{2}$As$_{2}$ (a) and BaFe$_{2}$As$_{2}$ (b) obtained within LDA 
(0~eV corresponds to the Fermi energy)}
\label{Ba_vs_K_DFT}
\end{figure}

{\bf Spectral properties.}~  
The orbitally resolved densities of states of KFe$_{2}$As$_{2}$ obtained 
within LDA are shown in the upper panel of Fig.\ref{Ba_vs_K_DFT} in 
comparison with the result obtained for BaFe$_{2}$As$_{2}$. In each case 
the Fe-$d$ states form a band with total width of $W\approx4$~eV located 
in the approximate interval $(-2,+2)$~eV. Therefore the on-site Coulomb 
parameter $U$ is comparable with the band width ($W/U\sim1$), implying 
that correlation effects are important. Both compounds have similar shape 
and relative positions of the spectral functions on the energy axis. 
However, in the case of KFe$_{2}$As$_{2}$ the Fermi level is located 
approximately 150~meV lower than in BaFe$_{2}$As$_{2}$ due to hole doping.     

\begin{figure}[t]
\centering \vspace{0.0mm}
\includegraphics[width=0.85\linewidth,angle=0]{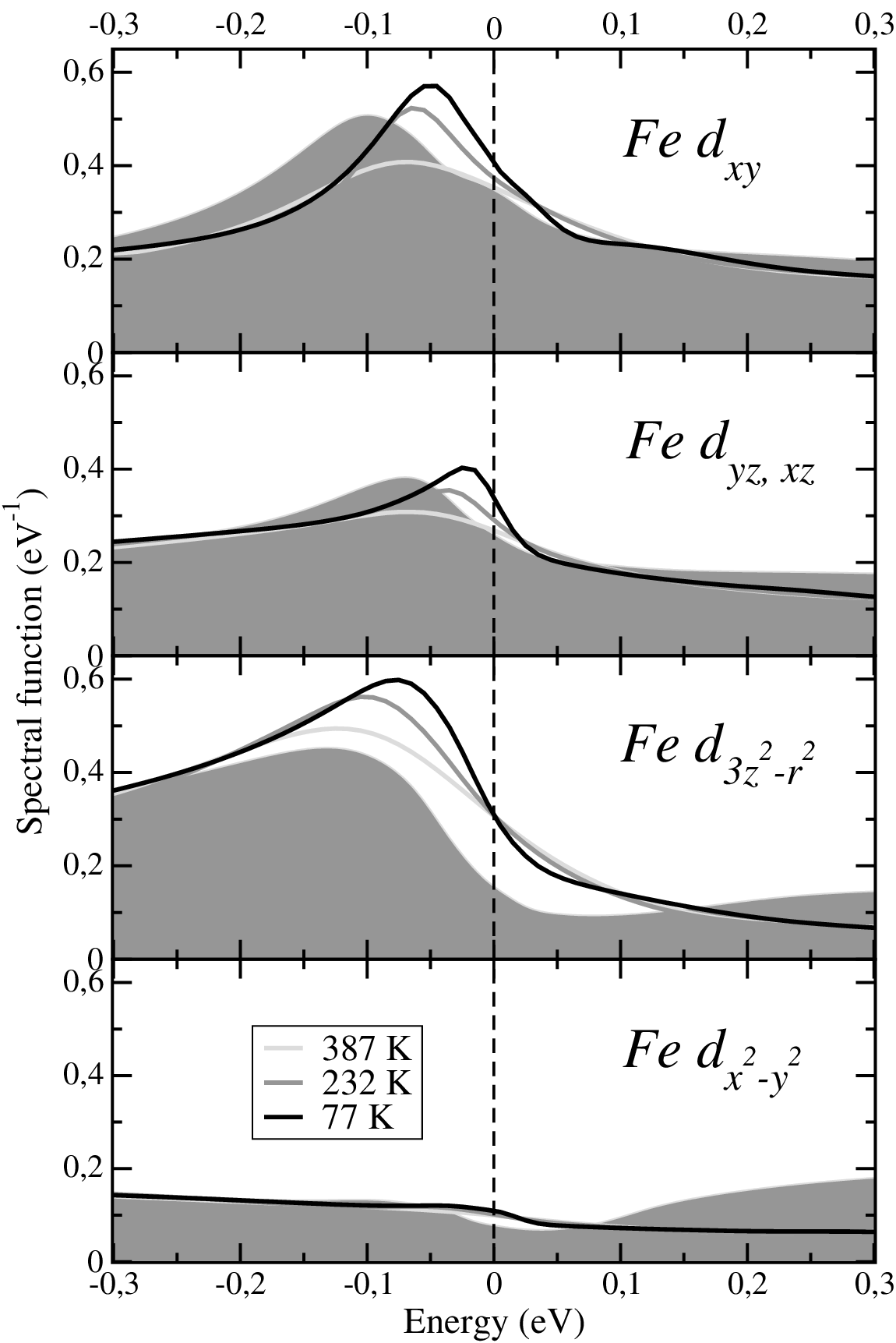}
\vspace{0.0mm}
\caption{Fig.3.~Temperature evolution of Fe-{\it d} spectral functions 
of KFe$_{2}$As$_{2}$ computed by LDA+DMFT. The spectral functions of 
BaFe$_2$As$_{2}$ corresponding to the temperature $T$=232~K taken 
from Ref.~\cite{SkornyakovBaFe2As2} are shown for comparison as shaded 
areas. The Fermi energy is set to 0~eV}
\label{DMFT_spectrum}
\end{figure}

The Fe-$d$ spectral functions of KFe$_{2}$As$_{2}$ computed within 
LDA+DMFT for the temperature window from 77 to 580~K are presented 
in Fig.\ref{DMFT_spectrum} along with the result for BaFe$_{2}$As$_{2}$
~\cite{SkornyakovBaFe2As2}. As in other pnictide superconductors the 
dynamical Coulomb correlations renormalize the spectrum in the vicinity 
of the Fermi energy and smear some fine details observed within the 
LDA, but the overall shape of the curves remains unchanged. This 
renormalization reduces the distance between the peaks in the Fe-$d$ 
spectral functions and the Fermi energy. In particular, the peak in 
the $yz(xz)$ spectral function is now significantly closer to the 
Fermi level compared to that in BaFe$_{2}$As$_{2}$.  

Quantitatively, the strength of the electronic correlations can be 
estimated by the increase of the effective masses in comparison with 
the LDA results. In the case of a single orbital the mass renormalization 
is expressed by the derivative of the self-energy $\Sigma(\omega)$ as 
$m^{*}/m_{b}=(1-\partial\mathrm{Re}\Sigma(\omega)/\partial\omega)$, 
where $m^{*}$ denotes the effective mass in LDA+DMFT and $m_{b}$ is 
the band mass obtained in LDA. In our calculation the self-energy is 
a diagonal matrix which leads to an orbital dependence of the masses. 
The calculated values of $m^{*}/m_{b}$ for every Fe-$d$ orbital are 
shown in Table.\ref{Table1}. The largest mass renormalization, 4.47, 
corresponds to the $xy$ orbital. Electronic correlations in the other 
$d$-orbitals are weaker with $m^{*}/m_{b}$ ranging from 2.22 to 4.02. 
The computed values of $m^{*}/m_{b}$ are in good agreement with previous 
theoretical estimations~\cite{Hardy} as well as with the ARPES data of 
Yoshida et~al.~\cite{Yoshida}. The result that the electrons in 
the $\left|xy\right>$-derived bands are the most correlated ones 
followed by the $\left| yz\right>$, $\left| 3z^2-r^2\right>$, and 
$\left| x^2-y^2\right>$ states, is in qualitative agreement with the 
conclusion on the proximity of KFe$_{2}$As$_{2}$ to an orbital-selective 
Mott transition reported by Hardy et~al.~\cite{Hardy}. It should 
also be noted that the response of the electrons occupying states with 
larger $m^{*}/m_{b}$ is more Curie-Weiss-like, while the temperature 
dependence of the susceptibilities corresponding to the other orbitals 
is less prononunced. A similar result was obtained in Ref.~\cite{HK} 
for the local susceptibility of LaFeAsO.

\begin{table}
\caption{Table~I.~The effective mass enhancement $m^{*}/m_{b}$ for different orbitals 
of the {\it d} shell}
\begin{tabular}{c|cccc}
\hline
\hline
           & d$_{xy}$ & d$_{yz}$ & d$_{3z^{2}-r^{2}}$ & d$_{x^{2}-y^{2}}$ \\
\hline
$m^{*}/m$  & 4.74     & 4.02     & 2.96               & 2.22              \\
\hline
\hline
\end{tabular}
\label{Table1}
\end{table}


{\bf Discussion.}
To explain why the magnetic susceptibility of KFe$_{2}$As$_{2}$ behaves 
qualitatively different from that of the other iron pnictides it is 
instructive to compare the spectral properties of KFe$_{2}$As$_{2}$ 
and the other end member of the '122' family, BaFe$_{2}$As$_{2}$. As 
noted above, in KFe$_{2}$As$_{2}$ the Fermi energy is lower than in 
BaFe$_{2}$As$_{2}$ due to hole doping. Already on the level of LDA this 
results in a smaller distance between the peaks of the Fe-$d$ spectral 
functions and the Fermi energy. Since correlation effects in 
KFe$_{2}$As$_{2}$ are stronger than in BaFe$_{2}$As$_{2}$ the peaks 
obtained within LDA+DMFT come to lie even closer to the Fermi energy. 
This is clearly seen in Fig.\ref{DMFT_spectrum} where the orbitally 
resolved Fe-$d$ spectral functions of BaFe$_{2}$As$_{2}$ are shown for 
comparison. 

In our previous study~\cite{SkornyakovBaFe2As2} we investigated the 
temperature behavior of the magnetic susceptibility in a model where 
the spectral function has a sharp peak below the Fermi energy. It was 
shown that the behavior of the magnetic susceptibility is determined 
by the thermal excitations corresponding to the states forming the 
peak, and that the distance between the peak and the Fermi energy can 
be regarded as a parameter controlling the magnetic response of the 
system. According to the present LDA+DMFT study, in KFe$_{2}$As$_{2}$ 
the peaks of the Fe-$d$ spectral functions are significantly closer 
to the Fermi energy than in BaFe$_{2}$As$_{2}$. Physically this means 
that the excitation of the states forming the peaks in KFe$_{2}$As$_{2}$ 
requires less energy than in BaFe$_{2}$As$_{2}$. An analysis of the 
model shows that the doping of 0.5 holes per iron atom is not sufficient 
to switch the system to the regime where the linear in $T$ behavior 
of the susceptibility no longer exists. Therefore we expect the linear 
increase of $\chi(T)$ of KFe$_{2}$As$_{2}$ to start at a lower 
temperature than in BaFe$_{2}$As$_{2}$, while the overall shape of the 
susceptibility curves is similar. Indeed, the decreasing part of the 
experimentally measured susceptibility of KFe$_{2}$As$_{2}$ is well 
described by our result.

We were not able to perform the susceptibility calculations for 
temperatures lower than 77~K because Monte-Carlo simulations become 
extremely time consuming. As a consequence the maximum of the 
susceptibility in KFe$_{2}$As$_{2}$ is not captured by our calculations. 
In our previous investigation we showed that the increasing part of the 
curve below that maximum can be interpreted as a quasilinear region 
in the vicinity of a turning point. In KFe$_{2}$As$_{2}$ a similar 
region is experimentally observed in the temperature window from 30 
to 80~K. It remains to be seen whether this low-temperature behavior 
can be explicitly reproduced in future LDA+DMFT studies.


In conclusion, the temperature dependence of the uniform magnetic 
susceptibility of KFe$_{2}$As$_{2}$ was investigated within the LDA+DMFT 
method. The temperature decrease of the computed susceptibility between 
125~K to 300~K agrees well with experiment. We found that, similar to other 
pnictides including the isostructural parent compound BaFe$_{2}$As$_{2}$, 
the Fe-$d$ spectral functions of KFe$_{2}$As$_{2}$ show sharp peaks 
below the Fermi energy. However, these peaks lie significantly closer 
to the Fermi level than in BaFe$_{2}$As$_{2}$. Making use of the 
scenario developed in our previous study of the magnetic properties 
of iron pnictides, we conclude that the qualitative difference between 
the magnetic susceptibilities of the two isostrucural end members 
of the '122' family is due to the smaller separation between the 
Fe-$d$ spectral functions and the Fermi energy in KFe$_{2}$As$_{2}$, 
which itself is a consequence of the effective hole doping and the 
stronger correlations in that compound. 


This work was supported by the Russian Foundation for Basic Research 
(Projects No.~13-02-00050-a, No.~13-03-00641-a) and the Ural Division 
of the Russian Academy of Science Presidium (Project No.~14-2-NP-164, 
No.~12-P2-1017). S.L.S. is grateful to the Dynasty Foundation for support. 
S.L.S and V.I.A. are grateful to the Center for Electronic Correlations 
and Magnetism, University of Augsburg for hospitality. This work was 
supported in part by the Deutsche Forschungsgemeinschaft through 
Transregio TRR 80. All results reported here were obtained using "Uran" 
supercomputer of IMM UB RAS.

\end{document}